\documentclass[fleqn,usenatbib]{mnras}

\usepackage[T1]{fontenc}

\DeclareRobustCommand{\VAN}[3]{#2}
\let\VANthebibliography\thebibliography
\def\thebibliography{\DeclareRobustCommand{\VAN}[3]{##3}\VANthebibliography}

\usepackage{graphicx} 
\usepackage{amsmath} 
\usepackage{amssymb} 
\usepackage{amsfonts}
\usepackage{soul,xcolor}
\usepackage{newtxtext,newtxmath}

\title[Short title, max. 45 characters]
{Hyperon bulk viscosity and $r$-modes of neutron stars}

\author[]{O. P. Jyothilakshmi$^1$, P. E. Sravan Krishnan$^1$, Prashant Thakur$^2$,\newauthor {V. Sreekanth$^1$\thanks{v\_sreekanth@cb.amrita.edu} and T. K. Jha$^2$} \thanks{{tkjha@goa.bits-pilani.ac.in}}
\\
$^1$Department of Sciences, Amrita School of Physical Sciences, Coimbatore, Amrita Vishwa Vidyapeetham, India\\
$^2$Department of Physics, BITS PILANI K K Birla Goa Campus, Goa 403726, INDIA}
\date{Accepted XXX. Received YYY; in original form ZZZ}

\pubyear{2022}

\date{Accepted XXX. Received YYY; in original form ZZZ}

\pubyear{2022}

\begin{document}
\label{firstpage}
\pagerange{\pageref{firstpage}--\pageref{lastpage}}
\maketitle
\begin{abstract}
We propose and apply a new parameterization of the modified chiral effective model to study rotating neutron stars with hyperon cores in the framework of the relativistic mean-field theory. The inclusion of mesonic cross couplings in the model has improved the density content of the symmetry energy slope parameters, which are in agreement with the findings from recent terrestrial experiments. The bulk viscosity of the hyperonic medium is analyzed to investigate its role in the suppression of gravitationally driven $r$-modes. The hyperonic bulk viscosity coefficient caused by non-leptonic weak interactions and the corresponding damping timescales are calculated and the $r$-mode instability windows are obtained. The present model predicts a significant reduction of the unstable region due to a more effective damping of oscillations. We find that from $\sim 10^8$ K to $\sim 10^{9}$ K, hyperonic bulk viscosity completely suppresses the $r$-modes leading to a stable region between the instability windows. Our analysis indicates that the instability can reduce the angular velocity of the star up to $\sim$0.3~$\Omega_K$, where $\Omega_K$ is the Kepler frequency of the star.
\end{abstract}




\section{Introduction}
The ever growing wealth of information from the observed gravitational wave (GW) events such as the GW190814 \citep{Abbott2020}, or the event GW170817 \citep{Abbott2017} points to the possibility of massive stars and hence to the role of exotic matter in compact stars. Over the last decade, the observations of high mass pulsars such as PSR J0740 + 6620 with a mass $2.14^{+0.10}_{-0.09} M_{\odot}$ \citep{Cromartie2019}, PSR ~J1614-2230 ($M = 1.928 \pm~ 0.04 M_{\odot}$) \citep{Fonseca2016} and PSR~ J0348 - 0432 ($M = 2.01 \pm~ 0.04~ M_{\odot}$) \citep{Antoniadis2013} severely constrains the equation of state (EoS) of neutron star matter, particularly when one considers exotics like hyperons or quarks in their inner shells. Their presence is known to have a softening effect on the EoS on one hand, while on the other hand we require a stiffer EoS in order to achieve a massive neutron star configuration. This is known as the 'hyperon puzzle' and a recent review of it can be found in \cite{Vidana2018}. In order to obtain high mass stars, one may consider the effect of repulsive hyperon-hyperon interaction \citep{Bednarek2012} or repulsive hyperonic three-body forces \citep{Vid00,Yam14} or even phase transition to deconfined quark matter \citep{Wei2019,Jha18}. Alternately, there also exists modified or extended theories of gravity \citep{Brax2008,Capozziello2016} which have been successfully applied to obtain high mass stars. 
Dense matter interactions in the EoS need to agree with astrophysical observations as well as various terrestrial experiments, at both low and high densities. 
Agreement with various terrestrial experiments such as the heavy ion collision (HIC) flow data \citep{Dan2002} or symmetry energy analysis \citep{Dan2009} or experiments for determining neutron skin thickness \citep{Brown2001, Horo2001} etc. enhances the reliability of the underlying EoS, which may be applied to various aspects of nuclear matter studies.

Studying the dynamical properties of rotating neutron stars is a domain which brings out various interesting features when one assumes a perfect fluid. It is known that the centrifugal force of a rotating star counters gravitational force and hence one can expect massive stars to be fast rotors, at least in the initial stages of the stellar evolution. As a result of rotation a star may experience damping due to unstable oscillations such as the $r$-modes. The $r$-modes are one of many pulsating modes that exist in neutron stars, and are characterized by the Coriolis force acting as the restoring force \citep{Andersson98}. The $r$-modes are unstable to emission of gravitational radiation (GR) by the Chandrashekhar-Friedman-Schutz (CFS) mechanism \citep{CFS, Friedman78}. It was shown in \cite{Andersson98} that the $r$-modes are unstable for all rotating perfect fluid stars irrespective of their frequency. Dissipative effects such as shear and bulk viscosities work towards suppressing GR driven instabilities and has been studied by various authors over the past few years \citep{Jones2001,Drago05,LO2002,Mohit06,Lind98,Dal04,Jaikumar2008,Jha10,Ofengeim2019} under various considerations. If the GR timescale is shorter than the damping timescale due to such dissipative processes, then the $r$-mode will be unstable and a rapidly rotating neutron star could lose a significant fraction of its rotational energy through GR. At higher temperatures ($T>10^9$ K), the dominant dissipation is due to bulk viscosity, which arises due to density and pressure perturbations, a consequence of the star being driven out of equilibrium by oscillations. The system tries to restore equilibrium through various internal processes. In the case of $r$-modes, since the typical frequencies are of the order of the rotational frequencies of the stars, the reactions that dominate are the weak processes. Within these weak processes, although the modified Urca processes involving leptons are important, it has been shown that non-leptonic processes involving hyperons contribute more significantly towards bulk viscosity at temperatures lower than a few times $10^9$ K \citep{LO2002}. Our goal here is to investigate the same using a chiral model calibrated to reproduce the desired nuclear matter properties, in particular the density content of the nuclear symmetry energy at both low and high densities.

In the following section, we present the attributes of the model that we consider for studying the pulsating modes. Subsequently we discuss briefly the hyperonic bulk viscosity and the $r$-mode damping phenomenon in Sections \ref{sec:bulk} and \ref{sec:rmode} respectively. In Section \ref{result}, we present our results and discussion. Throughout the paper we use $\hbar=c=1$.

\section{Formalism}
\label{Form}
\subsection{The Hadronic Model}
\label{model}
The effective Lagrangian of the model \citep{Malik2017} where nucleon doublet interacts through the exchange of the pseudo-scalar meson $\pi$, the scalar meson $\sigma$, the vector meson $\omega$ and the iso-vector $\rho-$meson is given by

\begin{eqnarray}
\label{lag}
{\cal L}&=& \bar\psi_{B}~\left[ \big(i\gamma_\mu\partial^\mu
         - g_{\omega}\gamma_\mu\omega^\mu
         - \frac{1}{2}g_{\rho}{\rho}_\mu\cdot{\tau}
            \gamma^\mu\big )\right]~ \psi_{B} -
            \nonumber\\
&&
          \bar\psi_{B}~\left[g_{\sigma~}~\big(\sigma + i\gamma_5
             \tau\cdot \pi \big)\right]~ \psi_{B} + \frac{1}{2}\big(\partial_\mu \pi\cdot\partial^\mu\pi
        + \partial_{\mu} \sigma \partial^{\mu} \sigma\big)
\nonumber \\
&&
        - \frac{\lambda}{4} \big(x^2 - x^2_0\big)^2
        - \frac{\lambda b}{6 m^2}\big(x^2 - x^2_0\big)^3 - \frac{\lambda c}{8 m^4}\big(x^2 - x^2_0\big)^4
        \nonumber \\
&&
        - \frac{1}{4} F_{\mu\nu} F_{\mu\nu}
        + \frac{1}{2}{g_{\omega B}}^{2}x^2 \omega_{\mu}\omega^{\mu}
        - \frac {1}{4}{ R}_{\mu\nu}\cdot{ R}^{\mu\nu}
        \nonumber \\
&&
        + \frac{1}{2}m^{\prime 2}_{\rho}{\vec \rho}_{\mu}\cdot{\vec \rho}^{\mu}
        +\eta_1\left(\frac{1}{2}g_\rho^2x^2\rho_\mu\cdot\rho^\mu\right).
\end{eqnarray}

The nucleon isospin doublet $\psi_B$ interacts with the aforesaid mesons. Following that we have the kinetic and the non-linear terms in the pseudo-scalar-isovector pion field $ \pi$, the scalar field $\sigma$, and higher order terms of the scalar field in terms of the chiral invariant combination of the two i.e., $x^2= ({\pi}^2+\sigma^{2})$. In the last two lines we have the field strength and the mass term for the vector field $\omega$ and the iso-vector field $ \rho$. The last term is the cross-coupling term between $\rho$ and $\sigma$ mesons with coupling parameter $\eta_1$. $g_{\sigma}, g_{\omega}$ and $g_{\rho}$ are the usual meson-nucleon coupling strengths of the scalar, vector and the iso-vector fields respectively. Here we shall be concerned only with the normal non-pion condensed state of matter. Hence we take $< \pi>=0$ and $m_{\pi} = 0$.

The spontaneous symmetry breaking (SSB) of the chiral symmetry renders the scalar field attaining the vacuum expectation value $x_0$. The masses of the nucleon ($m$), the scalar ($m_{\sigma}$) and the vector ($m_{\omega}$) mesons are then related to $x_0$ through
\begin{eqnarray}
m = g_{\sigma} x_0,~~ m_{\sigma} = \sqrt{2\lambda} x_0,~~
m_{\omega} = g_{\omega} x_0\ ,
\end{eqnarray}

\noindent
where $\lambda=(m_\sigma^2-m_\pi^2)/2f_\pi^2$ and $f_\pi$ is the pion decay constant which reflects the strength of SSB. Due to the cross-coupling between the $\sigma$ and $\rho$ mesons,
the mass of $\rho$-meson gets modified by the vacuum expectation value of $\sigma$ meson as 
$m_\rho^2=m^{\prime 2}_\rho+ \eta_1g_\rho^2 x_0^2$.

We recall here that in the relativistic mean field approach, the model has been applied to study matter at finite temperature and low density \citep{Jha2004}, and also the structure and composition of neutron stars \citep{Jha2006, Jha2007, Jha2008, Jha2008a}, but without the present modifications. 
The details of the mesonic field and related equations can be found in \cite{Malik2017}. The energy density and the pressure including the octet of baryons ($B = n^0, p^+, \Lambda^0, \Sigma^{+,0,-}, \Xi^{-,0}$) along with leptons ($L = e^-, \mu^-$) are
{\small
\begin{eqnarray}
\label{eosen}
      \epsilon &=& \frac{1}{\pi^2}\sum_{B}\int_0^{k_{F_B}}k^2\sqrt{k^2+m^*{^2}}dk
                       + \frac{m^2}{8 C_{\sigma B}}(1-Y^2)^2 \nonumber\\
      &&- \frac{b}{12 C_{\sigma B}C_{\omega B}}(1-Y^2)^3 + \frac{c}{16 m^2 C_{\sigma B}
            C_{\omega B}^2}(1-Y^2)^4  
         \nonumber \\
      &&+ \frac{1}{2}m_\rho^2\Big[1- \eta_1(1-Y^2) (C_{\rho B}/C_{\omega B})\Big] (\rho_{3B}^0)^2 \nonumber \\
      &&+ \frac{1}{2}m_{\omega}^2\omega_{0}^2Y^2 + \frac{1}{\pi^2}\sum_{L}\int_0^{k{F_L}}k^2\sqrt{k^2+m^*{^2}}dk, \\
\label{eospr}
   p &=& \frac{1}{3 \pi^2}\sum_{B}\int_0^{k_{F_B}}\frac{k^4}{\sqrt{k^2+m^*{^2}}}dk
         - \frac{m^2}{8 C_{\sigma B}}(1-Y^2)^2 \nonumber \\
     &&+ \frac{b}{12 C_{\sigma B}C_{\omega B}}(1-Y^2)^3 - \frac{c}{16 m^2
            C_{\sigma B}C_{\omega B}^2}(1-Y^2)^4     \nonumber\\
     && + \frac{1}{2}m_\rho^2 \Big[1- \eta_1(1-Y^2) (C_{\rho B}/C_{\omega B})\Big] (\rho_{3B}^0)^2
     \nonumber \\
     && + \frac{1}{2}m_{\omega}^2\omega_{0}^2Y^2 + \frac{1}{3 \pi^2}\sum_{L}\int_0^{k{F_L}}\frac{k^4}{\sqrt{k^2+m^*{^2}}}dk.
\end{eqnarray}}
The model parameters are listed in Table~\ref{tab:table1}, along with the corresponding nuclear matter saturation properties. { In the equations above $k_{F_B}$ is the Fermi momentum of each baryon species and $\gamma = 2$ is the spin degeneracy factor}. 
For a given baryon $B$,  $C_{\sigma B} \equiv g_{\sigma B}^2/m_\sigma^2$, $C_{\omega B}\equiv g_{\omega B}^2/m_\omega^2$ and $C_{\rho B} \equiv g_{\rho B}^2/m_\rho^2$ are the scalar, vector and iso-vector coupling parameters respectively, and $B$ and $C$ are higher order scalar couplings.
As emphasized in \citep{Malik2017}, the inclusion of the cross term ($\sigma-\rho$ with coupling $\eta_1$) has greatly improved the density content of the symmetry energy slope parameters.  For example, $L_0$ (the first derivative) decreased from 90 to 61 MeV while $J_1$, the symmetry energy defined at $\rho_1$= 0.1 fm$^{-3}$ increased from 22 to 24 MeV, at $J_0 = 32.5$ MeV. These values are in excellent agreement with the data obtained from various terrestrial experiments and astrophysical observations \citep{LI2013}. Moreover, the nuclear matter incompressibility value has dropped from a value of 300 MeV to 211 MeV in the present model, which previously was a bit too large in comparison to the recent experimental bounds \citep{Shlomo2006}. It would be interesting to compare the predictions of the modified model with the previous one. 

\begin{table}
   \caption{\label{tab:table1} The model parameters such as the meson-nucleon coupling constants $C_{\sigma N}, C_{\omega N}, C_{\rho N}$, and the higher order scalar field couplings $B$ and $C$ is in the top row along with $\eta_1$, the $\sigma-\rho$ mesonic cross coupling. The corresponding saturation properties such as the saturation density $\rho_0$, the nucleon effective mass, the nuclear matter incompressibility $K$, energy per particle $e_0$, symmetry energy $J_0$ and its slope $L_0$ for the model are enlisted.}
\begin{center}
\begin{tabular}{cccccc}  
\hline
$C_{\sigma N}$ & $C_{\omega N}$ & $C_{\rho N}$ & $\eta_1$ &  $B$ & $C$   \\ (fm$^2$) & (fm$^2$) & (fm$^2$) &  & (fm$^2$) & (fm$^4$) \\
\hline
8.81 & 2.16 & 13.00 & -0.85 & -12.08 & -36.44  \\
\hline
\\
$\rho_0$ & $m^{\star}/m$ & $K$ & $e_0$ & $J_0$  & $L_0$ \\
(fm$^{-3}$) &   & (MeV) & (MeV) & (MeV) & (MeV) \\
\hline
0.151 &  0.85  & 211  & -15.8 & 32.5  & 61  \\
\hline
\end{tabular}
\end{center}
\end{table}
\subsection{\label{sec:MR} Hyperon rich matter \& stellar profile}
For the inclusion of hyperons, their interaction strength or the couplings needs to be specified, which plays a pivotal role in the underlying EoS at higher densities. We fix them in conformity with the available data from hypernuclei experiments and define them with respect to the nucleon couplings, such as the scalar $x_{\sigma H} = g_{\sigma H}/g_{\sigma N}$, vector $x_{\omega H} = g_{\omega H}/g_{\omega N}$ and isovector $x_{\rho H} = g_{\rho H}/g_{\rho N}$ couplings for each hyperon species $H$. The binding energy of each species is reproduced at $\rho_0$ of symmetric nuclear matter by fixing the scalar coupling and calculating the corresponding vector counterpart by using the following equation
\begin{eqnarray}
    \left(\frac{B}{A}\right)_H = x_{\omega H}g_{\omega N} \omega_0 + m_H^* - m_H.
\end{eqnarray}
Here $m_H^* = m_H Y$ is the effective mass of a particular hyperon species. { The meson-hyperon couplings have been fixed to reproduce the hyperon binding energies in matter.} Recent experimental data indicates that the $\Lambda$ and $\Xi$ hyperons are bound with -28 MeV and -14 MeV respectively \citep{Gomes2015,Gal2016}. A recent work even estimates the potential depth as $U_{\Xi}\geq$ -20 MeV \citep{Gal2021}. Owing to the large uncertainty in the value of $U_\Sigma$ ($-10 \leq U_{\Sigma}^N \leq 50$ MeV) \citep{Gal2000}, we choose the potential depth of $U_{\Sigma} = 30$ MeV. In the present calculation we take $x_{\sigma H} = 0.7$ and $x_{\rho H} = x_{\omega H}$.

To calculate the EoS of charge neutral hyperon rich stellar matter, the condition of charge neutrality needs to be invoked which is
\begin{eqnarray}
\sum_{B}Q_{B}\rho_{B}+\sum_{l}Q_{l}\rho_{l}=0,
\end{eqnarray}
\noindent
where $\rho_{B}$ and $\rho_{l}$ are the baryon and the lepton (e,$\mu$) number densities, while $Q_{B}$ and $Q_{l}$ are their respective electric charges.
The density at which a particular hyperon species will appear depends on the chemical potentials of the neutron $\mu_n$ and the electron $\mu_e$. The chemical equilibrium condition of each baryon species $\mu_B$ is given by
\begin{eqnarray}\label{be}
    \mu_B = \mu_n - Q_B \mu_e,
\end{eqnarray}
where $Q_B$ is the electric charge of the concerned baryon species. 

The equations of the structure of a relativistic, spherical and static star composed
of a perfect fluid were derived from Einstein's equations by Tolman, Oppenheimer and Volkoff. These are known as Tolman-Oppenheimer-Volkoff equations (TOV) and are given as \citep{Oppen1939,Tolman1939}
\begin{equation}
    \frac{dP}{dr}=-\frac{G}{r}\frac{\left[\varepsilon+P\right]\left[M+4\pi r^3 P\right ]}{(r-2 GM)},\\
    \frac{dM}{dr}= 4\pi r^2 \varepsilon.
    \label{tov1}
\end{equation}
\noindent
Here $G$ is the gravitational constant and $M(r)$ is the enclosed gravitational mass.
For the specified EoS, they can be integrated from the origin as an initial value problem for a given choice of central energy density $(\varepsilon_c)$.
The value of $r~(=R)$, where the pressure vanishes defines the surface of the star.
We solve these equations to study the structural properties of a static neutron star using EoS as derived and given in Eq. (\ref{eosen}) and Eq. (\ref{eospr}) for the hyperon rich neutron star matter. To calculate the rotating properties, we use the $RNS$ code which is available online \citep{Steir95}.

\subsection{\label{sec:bulk} Hyperon Bulk Viscosity}
Bulk viscosity arises when a change in volume of a fluid element results in a release of heat and drives the system out of equilibrium. Various internal processes are then set up, which tend to restore equilibrium. These processes act on a characteristic timescale known as the relaxation timescale `$\tau$'. The damping due to bulk viscosity is quantified by the real part of the coefficient of bulk viscosity $\zeta$, which relates the perturbed pressure $p$ and the thermodynamic pressure $\Tilde{p}$ to the expansion of the fluid as \citep{Landau1987}
\begin{eqnarray}
p-\tilde {p} = -\zeta \mathbf{\nabla}\cdot{\mathbf v},
\end{eqnarray}
\noindent
where ${\mathbf v}$ is the velocity of the fluid element. A relativistic expression for { the real part of} $\zeta$ within a relaxation time approximation is given as \citep{Landau1987,LO2002}
\begin{eqnarray}
\textrm{Re}[\zeta] = \frac{p\left(\gamma_{\infty}-\gamma_{0}\right) \tau}{1+\left( \omega\tau\right)^2 }, \label{zeta}
\end{eqnarray}
where $\omega$ is the angular frequency of the perturbation in a co-rotating frame and $\tau$ is the net microscopic relaxation time. The terms $\gamma_{\infty}$ and $\gamma_0$ involving logarithmic derivative of pressure with respect to baryon density are 
defined as
\begin{eqnarray*}
 \gamma_{\infty} &=& \frac{n}{p}\left(\frac{\partial p}{\partial n}\right)_x,\\
 \gamma_0 &=& \frac{n}{p}\left[\left(\frac{\partial p}{\partial n}\right)_x + \left(\frac{\partial p}{\partial x}\right)_n \frac{d \tilde{x}}{dn}\right], 
\end{eqnarray*} 
where $x$ is a fluid variable, $n$ is the perturbed particle number density and $\tilde{x}$ is the thermodynamics state. The expression for $\gamma_{\infty}-\gamma_{0}$ then becomes
\begin{eqnarray}
\gamma_{\infty}-\gamma_{0} = -\frac{{\rho_B}^2}{p}
\frac{\partial p}{\partial \rho_n}\frac{d{\tilde{x}}_n}{d\rho_B} .
\label{gamma}
\end{eqnarray}
Here ${\tilde{x}}_n=\rho_n/\rho_B$ is the neutron fraction where $\rho_n$ is the neutron number density and $\rho_B$ is the total baryon number density.

\nocite{*}
\indent We now consider non-leptonic reactions involving hyperons which has been shown to give rise to high values of bulk viscosity in neutron star matter for temperatures in the range ($10^9 - 10^{10}$) K \citep{LO2002,Jones2001}.
These reactions are
\begin{eqnarray}
 n+n &\longleftrightarrow& p+\Sigma^-, \label{r1} \\
 n+p &\longleftrightarrow& p+\Lambda^0,\label{r2} \\
 n+n &\longleftrightarrow& n+\Lambda^0.\label{r3}
\end{eqnarray}
The rates of these reactions can be calculated from the tree-level Feynman diagrams involving the exchange of a $W$ boson. The reaction (\ref{r3}) is not considered, since it does not have a simple $W$-exchange picture. Its rate is an order of magnitude higher than that of (\ref{r2}). Consequently, the contribution to bulk viscosity is smaller
in the limit $\omega\tau\ll 1$
\citep{Jones2001}.
The expression for relaxation timescale { at temperature $T$} in the presence of both  $\Sigma^{-}$ and $\Lambda^0$ hyperons is given by \citep{LO2002,Mohit06},
\begin{eqnarray}
 \frac{1}{\tau}=\frac{(k_BT)^2}{192{\pi^3}}
\left(k_{\Sigma}\left\langle\vert\mathcal M_{\Sigma}^2\vert  \right\rangle
 + k_{\Lambda}\left\langle\vert\mathcal M_{\Lambda}^2\vert  \right\rangle\right) \frac{\delta \mu}{\rho_B \delta x_{n}}, \label{tau}
\end{eqnarray}
where $k_B$ is Boltzmann's constant, $k_{\Lambda}$ and $k_{\Sigma}$ are the Fermi momenta of ${\Lambda}$ and ${\Sigma}$ hyperons, $\delta \mu$ is the chemical potential imbalance, $\delta x_{n}$ is a small change between the perturbed and equilibrium values of the neutron fraction and $\left\langle\vert\mathcal M^2\vert  \right\rangle$ are the angle averaged, squared and summed over initial spinors matrix elements of the reactions calculated from the Feynman diagrams.
{ The factor $\delta \mu / \rho_B \delta x_{n}$ is determined by imposing conditions of electric charge neutrality and baryon number conservation, together with the condition that the non-leptonic strong interaction reaction $n+\Lambda^0 \longleftrightarrow p+\Sigma^- \label{r4}$
is in equilibrium, compared to weak interaction processes giving rise to bulk viscosity. For a region where both $\Lambda^0 $ and $\Sigma^-$ hyperons are present \citep{LO2002,Mohit06}
\begin{eqnarray}\label{delmuBoth}
\frac{\delta \mu}{\rho_B \delta x_{n}} &=& \alpha_{nn} +
\frac{(\beta_n-\beta_{\Lambda})(\alpha_{np}-\alpha_{\Lambda p}
+\alpha_{n\Sigma}-\alpha_{\Lambda\Sigma})}{2\beta_{\Lambda}
-\beta_{p}-\beta_{\Sigma}}
\nonumber \\
&& -\alpha_{\Lambda n} - \frac{(2\beta_n-\beta_p-\beta_{\Sigma})
(\alpha_{n\Lambda}-\alpha_{\Lambda\Lambda})}{2\beta_{\Lambda}
-\beta_{p}-\beta_{\Sigma}}.
\end{eqnarray}
}For a region where only $\Lambda^0$ hyperons are present \citep{Chatterjee2006}
\begin{eqnarray}
\label{delmuLambda}
\frac{\delta\mu}{\rho_B\delta x_n}&=\alpha_{nn}-\alpha_{\Lambda n}-\alpha_{n\Lambda}+\alpha_{\Lambda\Lambda}.
\end{eqnarray}
{ The quantities $\alpha_{ij}$ and $\beta_i$ are defined as $\alpha_{ij}=\left(\frac{\partial\mu_i}{\partial \rho_j} \right)_{\rho_{k},k\neq j}$ and $\beta_i = \alpha_{ni} + \alpha_{\Lambda i}-\alpha_{pi}-\alpha_{\Sigma i}$, where $\mu_i$ is the chemical potential of baryon $i$ given as
\begin{eqnarray*}
\mu_i&=\left(k_{Fi}^2+m_i^{*2}\right)^{\frac{1}{2}}+g_{\omega i}\omega_0+ \frac{1}{2}I_{3i}g_{\rho i}\rho_{03}.
\end{eqnarray*}
The above equation upon differentiation gives }
 \begin{align}
     \alpha_{ij}=&\frac{m_i^*m_i}{\sqrt{k_{Fi}^2+m_i^{*2}}}\frac{\partial Y}{\partial \rho_j}+\frac{\pi^2\delta_{ij}}{k_{Fi}\sqrt{k_{Fi}^2+m_i^{*2}}}+\frac{g_{\omega i}}{g_{\omega j}}\frac{1}{(Yx_0)^2}\nonumber\\
    &+\frac{I_{3i}I_{3j} g_{\rho i}g_{\rho j}}{2 m_{\rho}^2 F_j} - \frac{2 g_{\omega i}x_{0}}{(Yx_{0})^3} \left(\sum_{B}
\frac{ \rho_B }{g_{\omega B}} \right) \frac{\partial Y}{\partial \rho_{j}} \nonumber \\
    &- \eta_1 I_{3i}g_{\rho i} Y \left(\sum_B\frac{I_{3B}g_{\rho B}C_{\rho B}\rho_B}{m_\rho^2 C_{\omega B} F_B^2}\right)\frac{\partial Y}{\partial \rho_j},
 \end{align}
 where $\delta_{ij}$ is the Kronecker delta and $F_B$ is defined as
\begin{eqnarray}
F_B \equiv \Big(1 - \eta_1 (1-Y^2)C_{\rho B}/C_{\omega B} \Big). \nonumber
\end{eqnarray}
From the scalar field equation \citep{Malik2017}, we obtain $\frac{\partial Y}{\partial \rho_j}$ as
\begin{eqnarray}
    \label{dauYSR}
    \frac{\partial Y}{\partial \rho_{j}}=&  \frac{1}{D}\Bigg[\frac{2 C_{\sigma j}C_{\omega j}\rho_j}{m_{j}^2Y^4}
    -\frac{C_{\sigma j}m^*_j}{m_j Y\sqrt{k_{F_{j}}^2+m_{j}^{*2}}}\\
    &+ \frac{2\eta_1 C_{\sigma j}C_{\rho j}I_{3j}^2g_{\rho j}^2\rho_j}{C_{\omega j}m_j^2 m_\rho^2 F_j^2}\Bigg],
\end{eqnarray}
where,
\begin{eqnarray}
    D&=&\sum_B\Bigg [ Y +\frac{2BY}{C_{\omega B}}(Y^2-1)  +
    \frac{3CY}{C_{\omega B}^2}(Y^2-1)^2  \nonumber \\
    && -\frac{C_{\sigma B}\rho_{SB}}{m_B Y^2} +\frac{C_{\sigma B}}{m_B Y} \frac{\gamma}{(2\pi)^3}
    \int^{k_{F_{B}}}_o d^3k \frac{k^2 m_B}{(k^2+m^{*2}_B)^{3/2}} \nonumber\\
    && +\frac{4 C_{\sigma B}C_{\omega B}\rho_B^2}{m_{B}^2Y^5} +\frac{4Y\eta_1^2
    C_{\rho B}^2C_{\sigma B} I_{3B}^2 g_{\rho B}^2\rho_B^2}{C_{\omega B}^2 m_B^2 m_\rho^2 F_B^3} \Bigg] .
\end{eqnarray}

{ The above equations reduce to that of \cite{Jha10} when $\sigma-\rho$ coupling is not considered ($\eta_1=0$).} 
We use these equations to obtain the damping timescales due to hyperonic bulk viscosity.

\subsection{\label{sec:rmode}$R$-mode Damping}
For $r$-modes, the Coriolis force acts as the restoring force. For a star rotating with angular frequency $\Omega$, the frequency of $r$-mode in a co-rotating frame $\omega$ and in an inertial frame $\omega_0$ are given upto first order in $\Omega$ as \citep{Provost1981}
\begin{eqnarray}
    \label{omega}
    \omega&=&\frac{2m\Omega}{l(l+1)}+O(\Omega^3),\\
    \omega_0&=&\omega-m\Omega=\left(\frac{2}{l(l+1)}-1\right)m\Omega.
\end{eqnarray}
For the $r$-modes, $l=m$ \citep{Papa1978}
. The mode that is most unstable to gravitational emission through the CFS mechanism is the $l=m=2$ mode. \\

Dissipative effects such as shear and bulk viscosities can damp out the instabilities if the timescales of these effects are comparable to the timescales of gravitational radiation. To study the evolution of $r$-modes in the presence of these effects, we use the definition of the overall $r$-mode timescale $\tau_r$ given in terms of the timescales $\tau_i$ of each process under consideration, namely hyperonic bulk viscosity (H), bulk viscosity due to Urca processes (U),
shear viscosity ($\eta$) and gravitational radiation (GR) \citep{LO2002,Lind98}
\begin{eqnarray}
{1\over\tau_r(\Omega,T)}={1\over\tau_{GR}(\Omega)}
+{1\over\tau_{H}(\Omega,T)}+{1\over\tau_{U}(\Omega,T)}
+{1\over\tau_{\eta}(\Omega,T)}
.\label{tauR}
\end{eqnarray}

The timescales of hyperonic bulk viscosity and Urca processes to the lowest order in $\Omega$ are given by \citep{Lind98,LO2002,Mohit06}
 \begin{align}
 \label{tauB,U}
 \frac{1}{\tau_{H,U}} &= \frac{4\pi}{690}\left(\frac{\Omega^2R^{l-4}}{\pi G \bar\rho}\right)^2 \left[\frac{\int_0^R \textrm{Re}[ \zeta_{H,U}(r)]\left[1+0.86\left(\frac{r}{R}\right)^2\right]r^8dr}{\int_0^R \rho (r) r^{2l+2}dr}\right].\nonumber\\
\end{align}
Here $\bar\rho$ is the mean density of the non-rotating star. $R$ and $\rho(r)$ denote the radius and the density profile of the slowly rotating star respectively. The hyperonic bulk viscosity Re[$\zeta_H$] is given by Eq. (\ref{zeta}), while the bulk viscosity associated with modified Urca processes in the limit $ {\omega\tau\gg 1}$ is given by \citep{Saw89}
\begin{eqnarray}
 \zeta_U=146~\rho(r)^2\omega^{-2}
\left[\frac{k_B T}{1 MeV}\right]^6 {~\rm gcm^{-1}s^{-1}},
\label{zetaU}
\end{eqnarray}
where $\rho$ is in units of $gcm^{-3}$ and $\omega$ is in units of $s^{-1}$. The timescale associated with shear viscosity is given by \citep{Lind98}
\begin{eqnarray}
\frac{1}{\tau_{\eta}} = \frac{(l-1)(2l+1)}{\int_0^R dr \rho(r) r^{2l+2}}
\int_0^R dr \eta r^{2l}.\label{taueta}
\end{eqnarray}
\noindent
Here $\eta$ is calculated from the prominent $nn$ scattering
and is given by \citep{Flowers1979,Cutler1987}
\begin{eqnarray}
 \eta = 2\times 10^{18}\rho_{15}^{9/4}T_9^{-2} {~\rm gcm^{-1}\ s^{-1}}, \label{Eta}
\end{eqnarray}
where $\rho_{15}=\rho/(10^{15}~{\rm g/cm^3})$ and $T_9=T/(10^9~{\rm K})$
are the dimensionless density and temperature respectively.
Finally, the time scale of gravitational radiation is given by \citep{Lind98}
\begin{eqnarray}
 \frac{1}{\tau_{GR}}=&-&\frac{32 \pi G \Omega^{2l+2}}{c^{2l+3}}
\frac{(l-1)^{2l}}{\lbrack(2l+1)!!\rbrack^2} \left( \frac{l+2}{l+1}
 \right) ^{2l+2}
\nonumber \\
&\times&
\int_0^R \rho(r) r^{2l+2} dr. \label{tauGR}
\end{eqnarray}
With time scales for each process calculated, we calculate the net $r$-mode time scale $\tau_r$ from Eq. (\ref{tauR}). The mode decays as $e^{-t/\tau_r}$, and therefore the mode is stable if $\tau_r>0$. From Eq. (\ref{tauGR})  it is clear that $\tau_{GR} < 0$, which indicates that GR grows the modes and drives them to instability, while $\tau_H$, $\tau_U$ and $\tau_\eta$ are positive which indicate that they dampen the mode.
We next solve for the equation 1/$\tau_r(\Omega_C,T)=0$ where $\Omega_C$ is defined as the critical angular velocity for a star with temperature $T$. A star with angular velocity greater than $\Omega_C$ will be unstable and subject to GR emission, while one with angular velocity less than $\Omega_C$ will be stable.
 
\section{Results and Discussion}
\label{result}
\begin{figure}
\includegraphics[width=9.4cm]{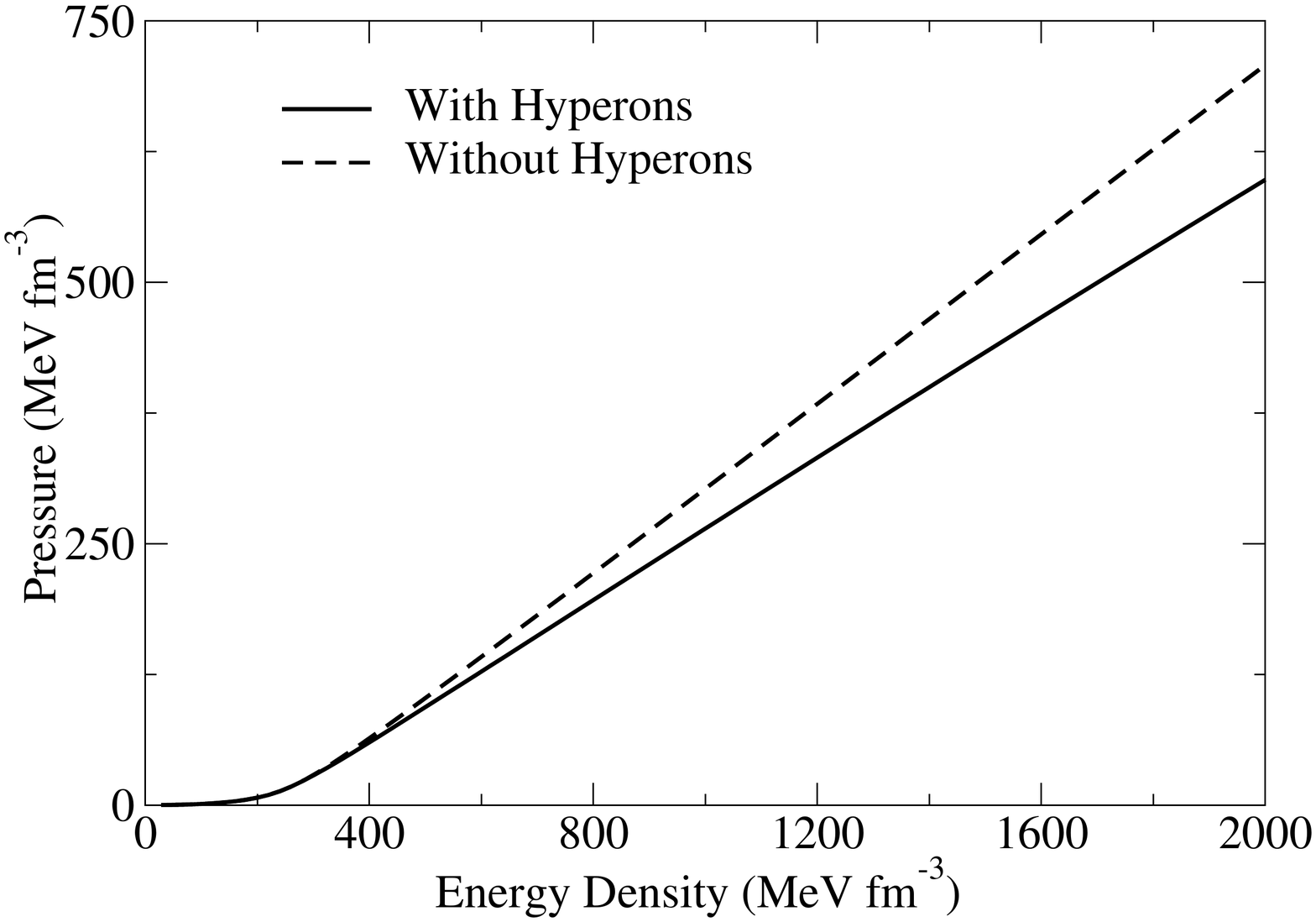}
\caption{\label{eos} Pressure as a function of energy density of neutron star matter with and without Hyperons.}
\end{figure}

The energy density and pressure of the charge neutral neutron star matter consisting of the octet of baryons is plotted in Fig. \ref{eos}, where the effect of hyperonization in the matter can be clearly seen on the underlying EoS. At $\approx 1.7\rho_0$, $\Lambda^0$ appears first followed by $\Sigma^-$ and as a result, the EoS becomes softer. For example, we witness a 5\% drop in the pressure at $\approx 3\rho_0$, while at $\sim$ $7\rho_0$ the drop is approximately 12\%.

\begin{figure}
\includegraphics[width=9.4cm]{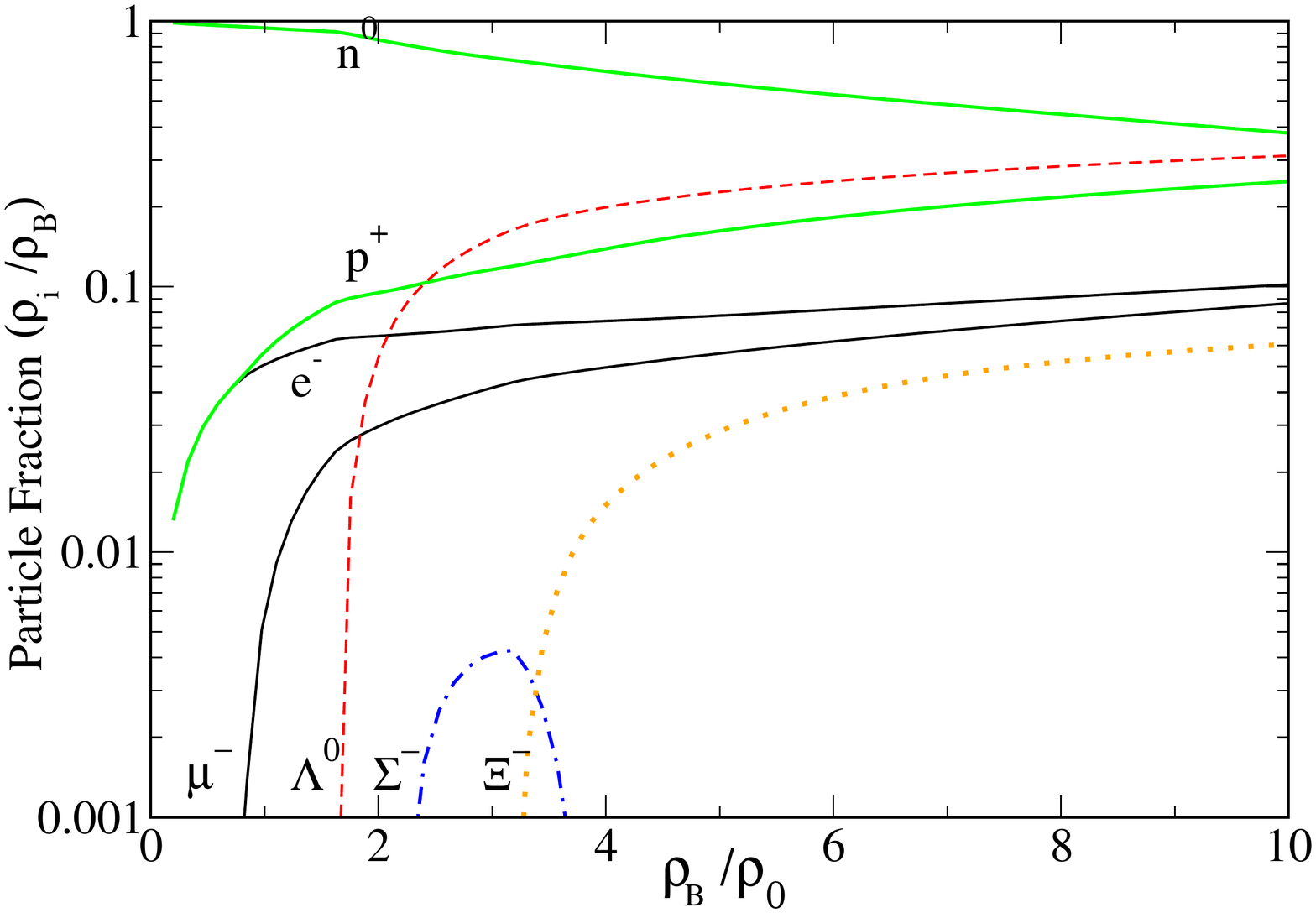}
\caption{\label{pf} Particle fraction of various octet of baryons (as marked in the figure) plotted as a function of normalized baryon number density.}
\end{figure}

The respective particle concentration is shown in Fig. \ref{pf}. Here we find that among the octet of baryons, only $\Lambda^0$ and $\Xi^-$ are present substantially in the neutron star matter. $\Lambda^0$ and $\Xi^-$ appear in the neutron star matter at $1.7 \rho_0$ and $3.3 \rho_0$ respectively and their concentration increases with increasing density. The model however does not predict the presence of any other hyperon species, though $\Sigma^-$ appears briefly in the matter and then disappears again. { The lepton number density is $\sim 17\%$ of the baryon number density in the star matter, which means that deleptonization of matter does not happen even at higher densities.}

\begin{figure}
\centering
\includegraphics[width=8cm,height=6.5cm]{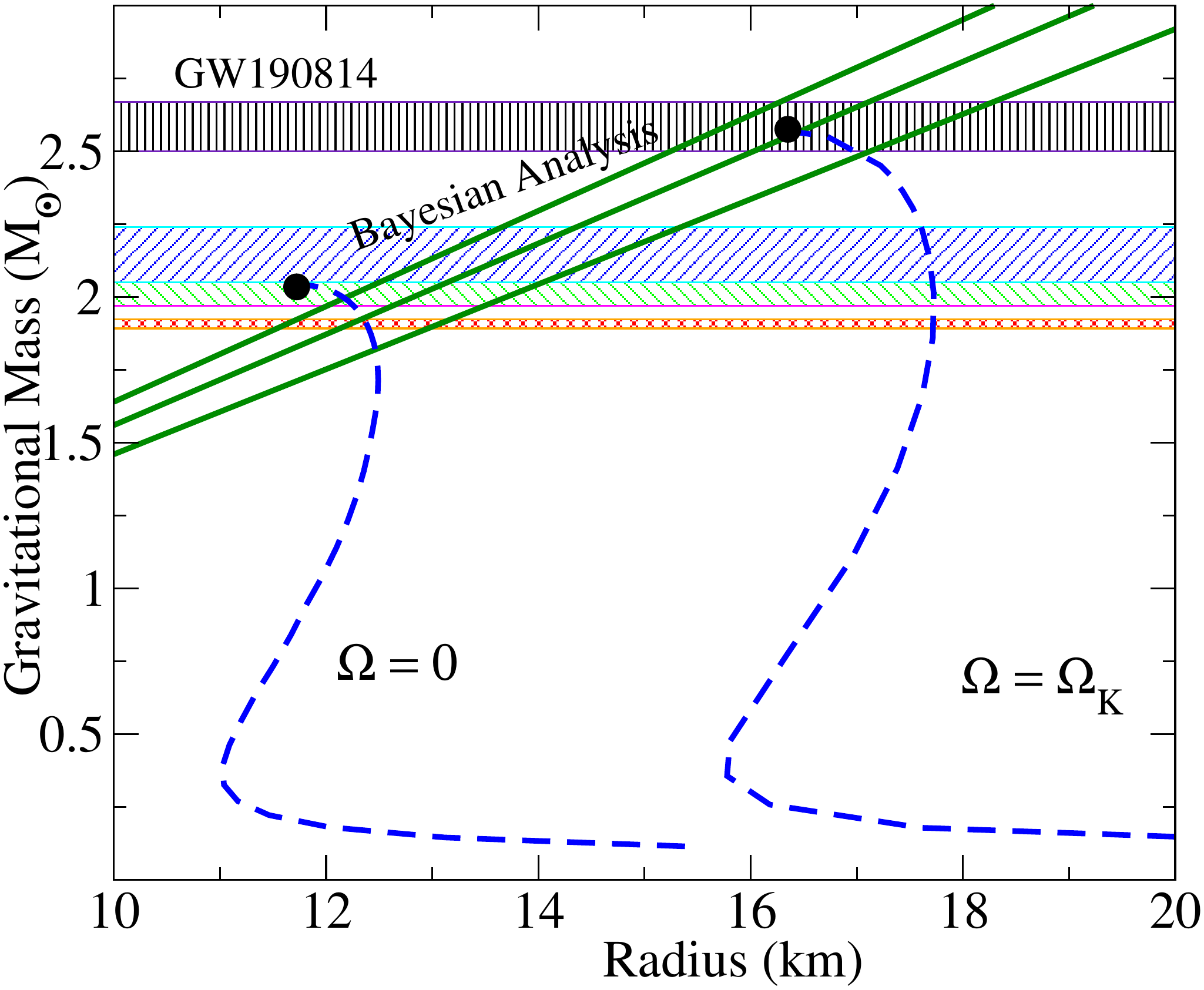}
\caption{\label{mr}Gravitational Mass plotted as a function of equatorial radius of the star for both static ($\Omega = 0$) and when it is rotating at Kepler's frequency ($\Omega = \Omega_K$) obtained using the RNS code, compared with the mass range of massive known pulsars (1.88 - 2.24) solar masses and the inferred mass range of the secondary component of the Gravitational wave `GW190814' data. The compactness parameter estimate from Bayesian analysis \citep{Riley_2019} is also indicated.}
\end{figure}

The global properties of the compact star are calculated for both the static case ($\Omega = 0$) and when it is rotating at the maximum possible limit ($\Omega = \Omega_K$) using the RNS code and is plotted in Fig. \ref{mr}. The difference in the latter is the additional centrifugal force, a component of which acts against gravity. In principle a massive star can potentially rotate faster at least in the initial stages of the stellar evolution. For our EoS, the maximum mass obtained for a non-rotating configuration is $2.05~M_{\odot}$, which occurs at the central density $\rho_c = 1.55 ~\times~10^{15}$ g cm$^{-3}$. The corresponding radius obtained is $11.86$ km. For the same central density, the Kepler frequency obtained is $\Omega_K = 1420$ Hz and the corresponding mass and equatorial radius are $2.57~M_{\odot}$ and $16.5$ km respectively. Our results agree well with some of the massive pulsars such as the millisecond pulsar ${ PSR}~ J0740+6620$ \citep{Cromartie2019}, $PSR~ J0348+0432$ \citep{Antoniadis2013} and $PSR~ J1614-2230$ \citep{Demo2010}. It is to be noted that the mass inference of the secondary component of the $GW190814$ ($\sim 2.6 ~M_{\odot}$) is yet to be identified as a pulsar or a black hole. However, this mass configuration may be attained by a rapidly rotating star, as  emphasized in recent findings  \citep{Most2020}. In such a scenario, it is very unlikely to be rotating anywhere close to break-up, since the spin-down time is typically much shorter than the time until the merger, and tidal forces do not significantly affect the rotation rate. At Kepler frequency, the compactness parameter obtained is $GM/Rc^2$ = 0.155. It may be noted that a recent work on Bayesian analysis to estimate the compactness parameter quotes a value $0.156^{+0.008}_{-0.010}$ \citep{Riley_2019}. However, this value corresponds to a much less massive and much slower neutron star.    

\begin{figure}
\includegraphics[width=8.5cm]{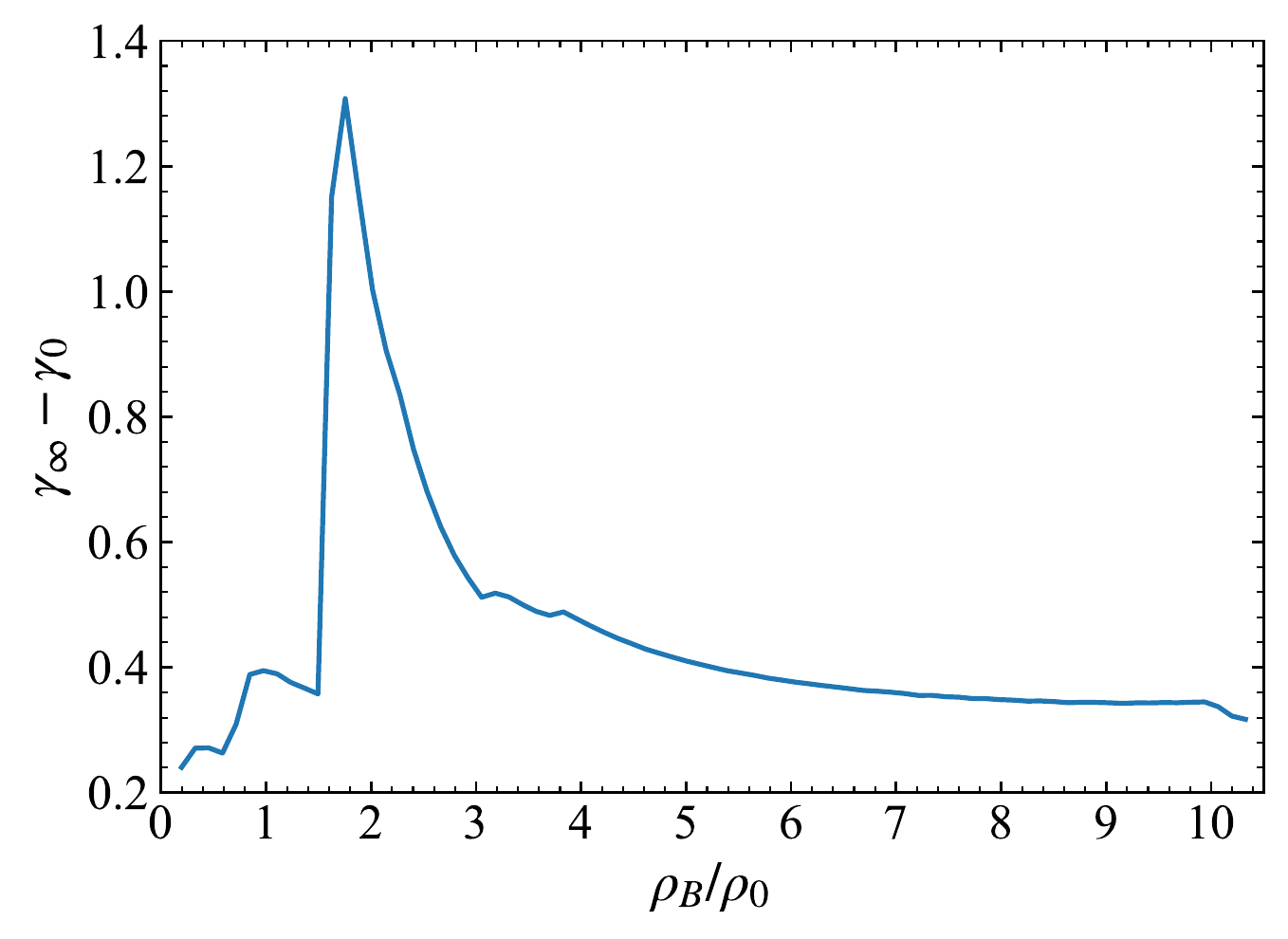}
\caption{\label{gammaplot} { Thermodynamic factor $\gamma_\infty - \gamma_0$ plotted against the normalized baryon number density.}}
\end{figure}

To calculate the coefficient of hyperonic bulk viscosity, the thermodynamic factor $\gamma_\infty - \gamma_0$ is calculated using Eq. \ref{gamma}, and is plotted against the normalized baryon number density $\rho_B/\rho_0$ in Fig. \ref{gammaplot}. The sudden rise in the value of the factor indicates the onset of hyperons in the matter. It is to be noted that $\Lambda$ is present for baryon densities for which $\rho_B/\rho_0>1.77$, while $\Sigma^-$ is present in the range $2.22<\rho_B/\rho_0<3.8$. To obtain the relaxation timescale $\tau$ (Eq. \ref{tau}), we consider the non-leptonic reactions given by Eqs. (\ref{r1}) and (\ref{r2}). The effective masses of the baryonic species and their Fermi momenta are used to calculate the matrix elements.
\begin{figure}
\includegraphics[width=8.5cm]{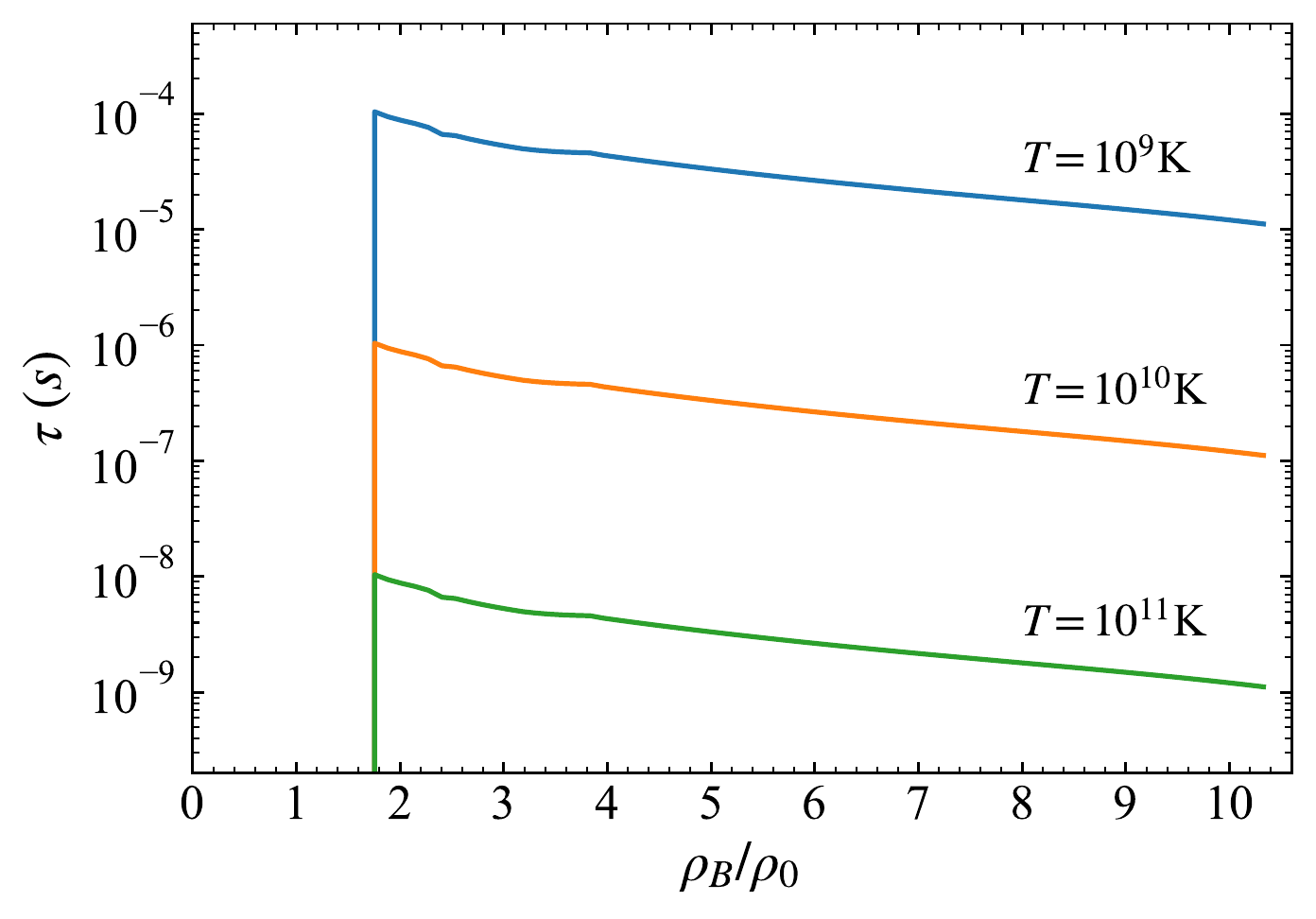}
\caption{\label{tauplot} The relaxation time scale $\tau$ plotted against normalized baryon number density $\rho_B/\rho_0$.}
\end{figure}
We use the axial-vector coupling values $g_{np} = -1.27,~g_{p\Lambda} = -0.72$
and $g_{n\Sigma}=0.34$ as measured in the $\beta$-decay of baryons at rest. 
The Fermi coupling constant is taken as $G_F = 1.166\times 10^{-11}$ MeV$^{-2}$ and sin$\theta_C=0.222$, where $\theta_C$ is the Cabibbo weak mixing angle \citep{LO2002}. The relaxation timescales (Eq. \ref{tau}) are calculated for various temperatures and is shown in Fig. \ref{tauplot}. The dependence of $\tau$ on the temperature goes as $T^{-2}$ and therefore the relaxation times are considerably larger for lower temperatures. At a given temperature, the relaxation time is found to decrease and then increase as the $\Sigma^-$ hyperons appear and disappear as evident from Eq. (\ref{tau}).\\

\begin{figure}
\includegraphics[width=8.5cm]{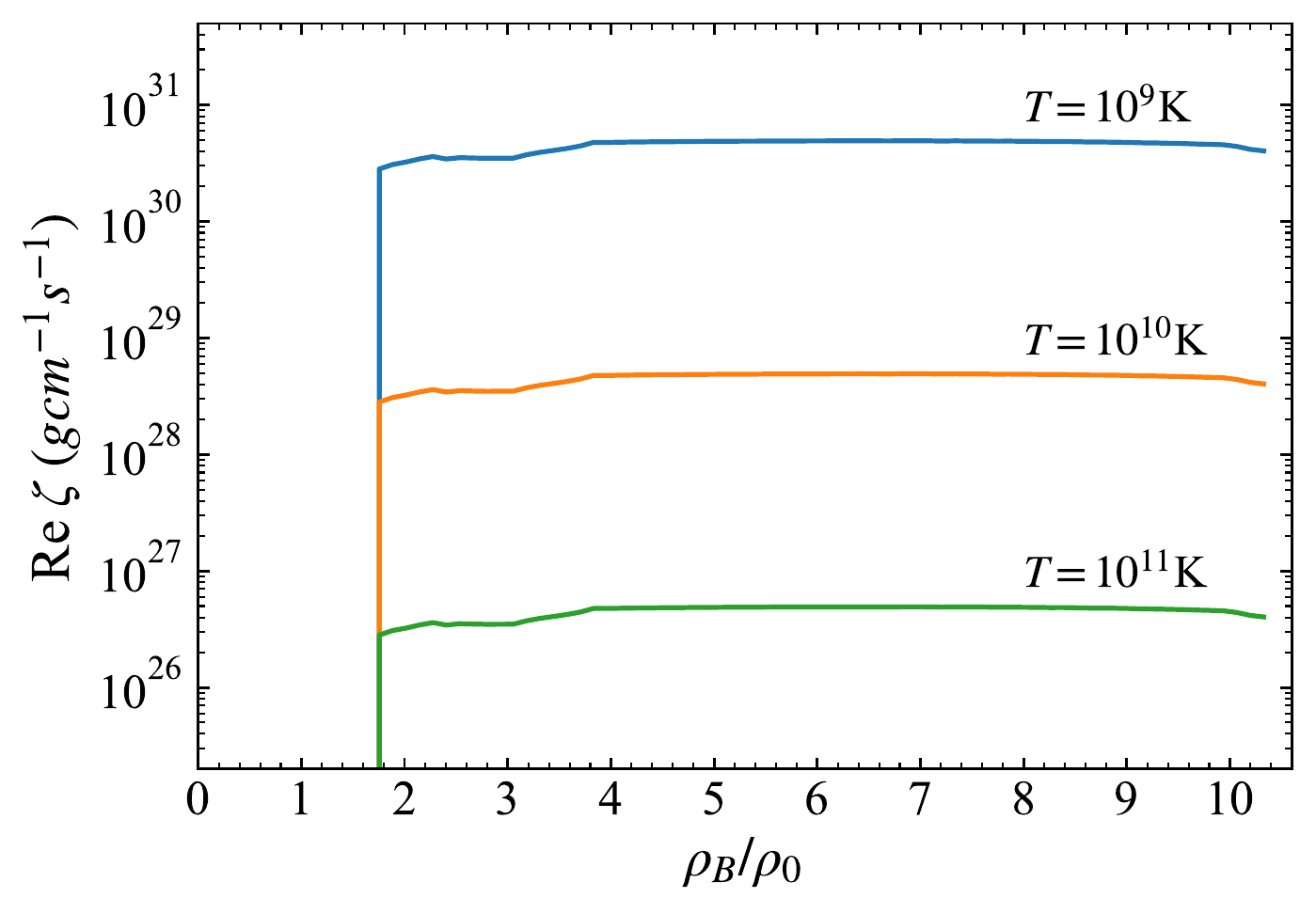}
\caption{\label{zetaplot}The hyperonic bulk viscosity coefficient Re[$\zeta$] plotted against normalized baryon number density $\rho_B/\rho_0$. }
\end{figure}

\begin{figure}
\includegraphics[width=8.5cm]{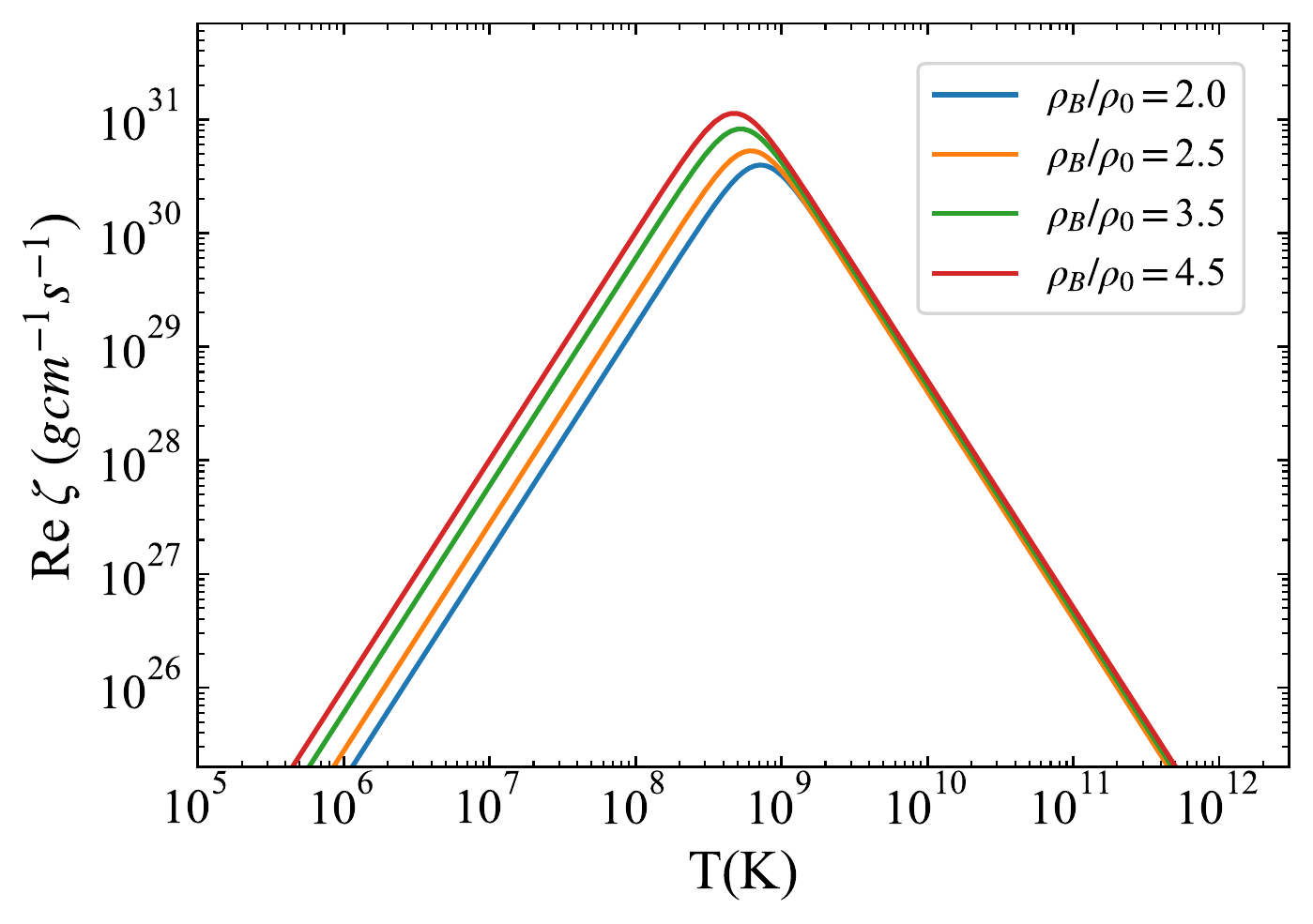}
\caption{\label{zetaVsT}The hyperonic bulk viscosity coefficient Re[$\zeta$] plotted against temperature $T$ for different normalized baryon number densities $\rho_B/\rho_0$.}
\end{figure}

\indent We then proceed to calculate the coefficient of hyperonic bulk viscosity from Eq. (\ref{zeta}). The $r$-mode frequency $\omega$ is obtained by setting $\Omega = \Omega_K$ in Eq. (\ref{omega}). The obtained values of Re[$\zeta$] are plotted against the normalized baryon density in Fig. \ref{zetaplot}. From Eq. (\ref{zeta}), it is clear that Re[$\zeta$] has a non-monotonic dependence on the relaxation timescale $\tau$ and therefore on the temperature $T$. In Fig. \ref{zetaVsT}, we plot Re[$\zeta$] against temperature for different baryon densities based on the presence of various hyperons as seen in Fig. \ref{pf}. We find that the bulk viscosity is maximum for temperatures around $10^8$ K for the relevant densities.
\begin{figure}
\includegraphics[width=8.5cm]{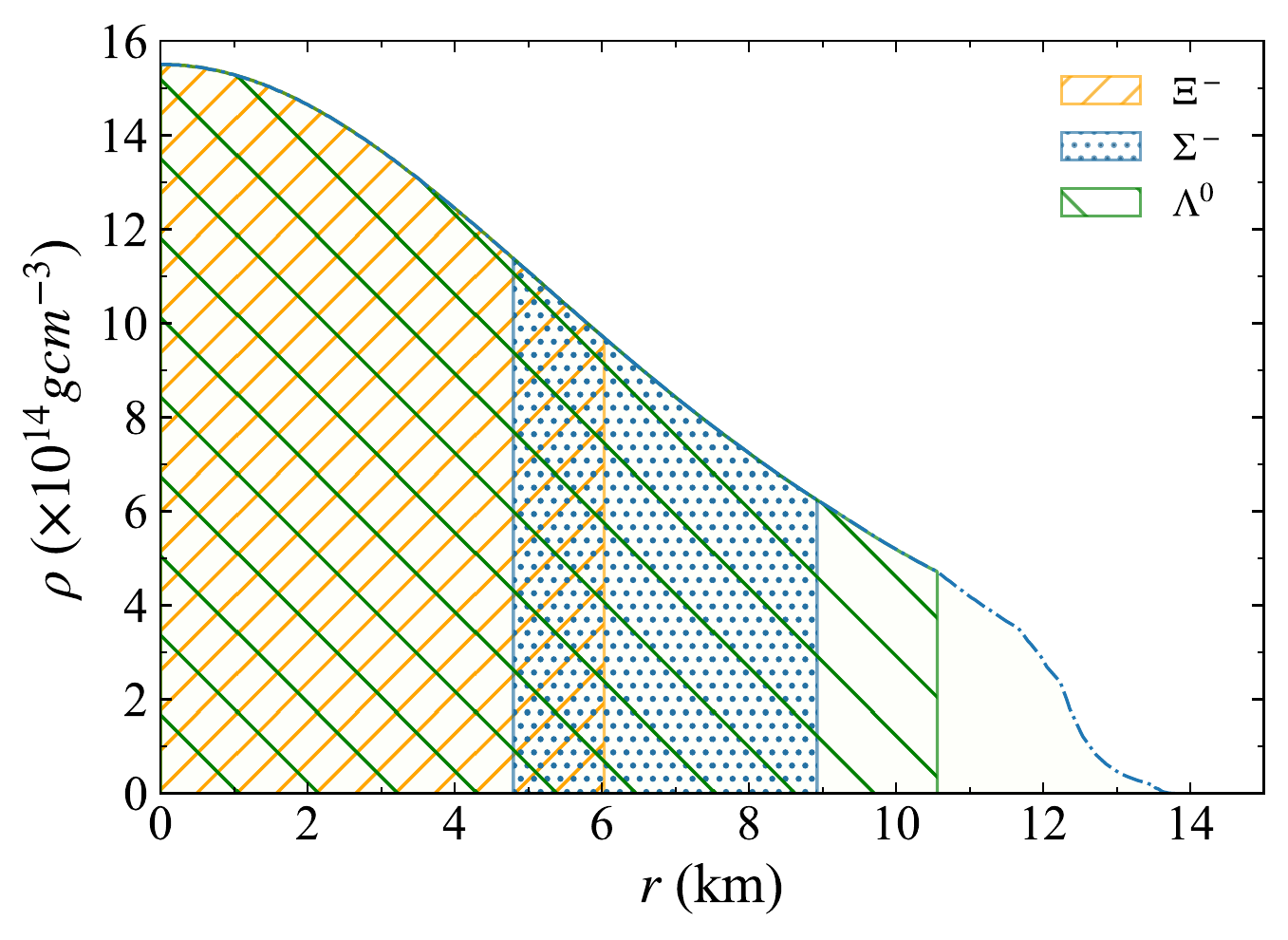}
\caption{\label{rhoplot} Density profile of the rotating star with threshold hyperon densities indicated.}
\end{figure}

\indent To investigate the effect of dissipation on the damping of $r$-modes, the timescales of each dissipative process must be calculated. To obtain the density profile $\rho(r)$ of the rotating star, the Hartle-Thorne slow rotation approximation \citep{Hart67, Hart68} is used, restricted to spherical deformations. For the configuration rotating at Kepler frequency, the mass and radius thus obtained are 2.39 $M_\odot$ and 14.07 km respectively. We plot $\rho(r)$ in Fig. \ref{rhoplot} with the range of densities corresponding to appearance of different hyperons. $\Lambda$ hyperons are found to be present up to a radius of 10.6 km, which corresponds to a density of 4.72 $\times~ 10^{14}$ g cm$^{-3}$; whereas $\Sigma^-$ hyperons are present between radii 4.38 km and 9.16 km, which corresponds to densities 1.14 $\times~ 10^{15}$ g cm$^{-3}$ and 5.99 $\times ~10^{14}$ g cm$^{-3}$ respectively. 

\indent To calculate the timescale due to hyperonic bulk viscosity $\tau_H$, we express the obtained value of Re[$\zeta$] as a function of $r$ using the equation of state and the density profile, i.e., Re[$\zeta](\rho(r))$ and calculate $\tau_H$ using Eq. (\ref{tauB,U}) for various temperatures. Similarly, the timescales due to Urca processes $\tau_U$, shear viscosity $\tau_\eta$ and gravitational radiation $\tau_{GR}$ are calculated using Eqs. (\ref{tauB,U}), (\ref{taueta}) and (\ref{tauGR}) respectively. With the obtained timescales, the net $r$-mode timescale $\tau_r$ is calculated using Eq. (\ref{tauR}). We then solve for the value of the critical angular velocity $\Omega_C$ defined for a particular temperature as
\begin{eqnarray}
 \frac{1}{\tau_r(\Omega_C,T)}=0.
\end{eqnarray}

\begin{figure}
\includegraphics[width=8.5cm]{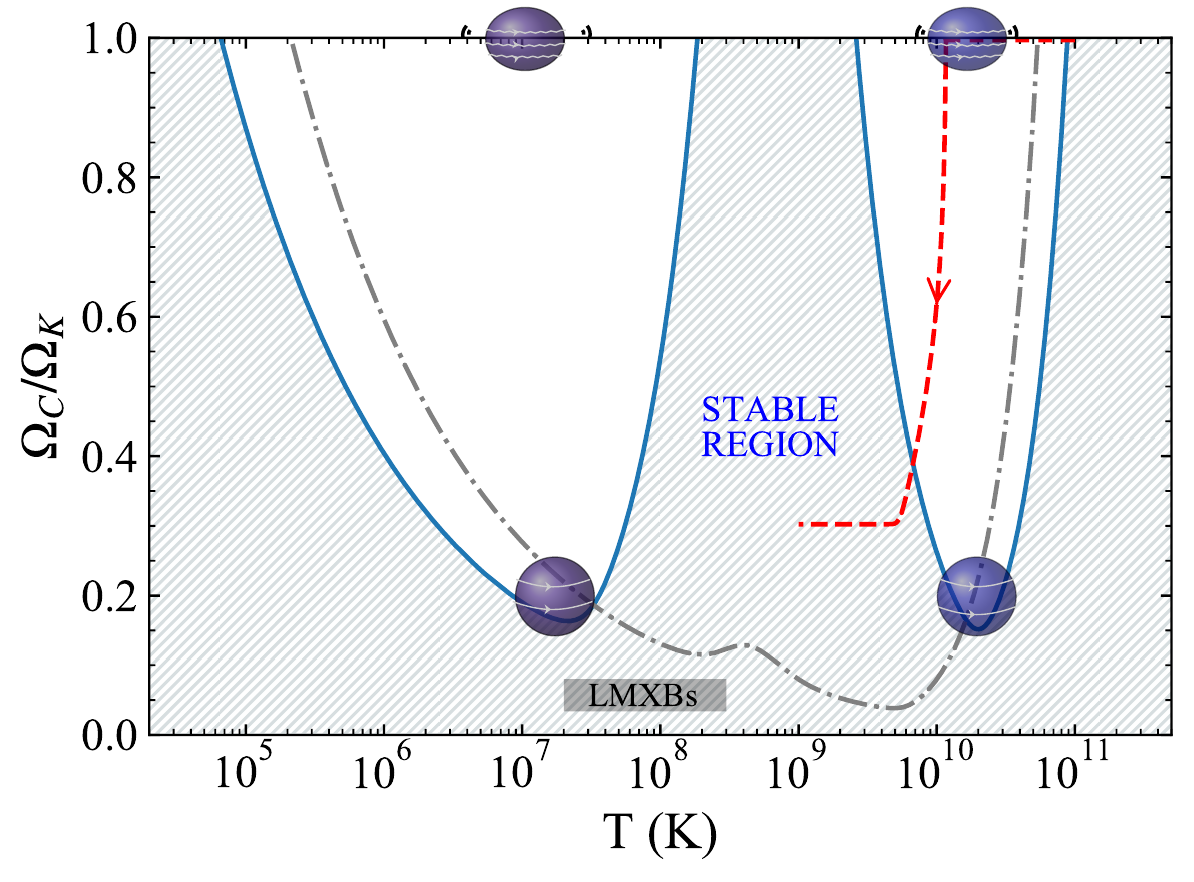}
\caption{Critical angular velocities normalized to Kepler frequency as a function of core temperature. The red dashed line shows the spin down of a neutron star due to $r$-modes for an initial amplitude of $10^{5}$, before getting stabilized by hyperon bulk viscosity. The shaded region denotes the observed Low Mass X-ray Binaries (LMXBs).
Also shown (in dotted lines) is the instability window when $\sigma-\rho$ coupling is not considered. \label{omegaplot}}
\end{figure}

The obtained values of $\Omega_C$ normalized to the Kepler frequency $\Omega_K$ is plotted against the corresponding temperatures in Fig. \ref{omegaplot}. { In the present work,  there exists two instability windows, one in the low temperature region and another one in the high temperature region, separated by a region of no instability formed by the complete damping of the $r$-modes due to hyperonic bulk viscosity. 
The low-temperature window falls between $6.9\times 10^4$ K and $1.8\times 10^8$ K while the high temperature window falls between $2.67\times 10^9$ K and $8.44\times 10^{10}$ K. As stated earlier, the hyperonic bulk viscosity reduces at lower temperatures and this results in the first instability window. The minima of the two windows occur at $0.16$ $\Omega_K$ and $0.15$ $\Omega_K$, which corresponds to temperatures of $2.09\times10^{7}$ K and $2.07\times10^{10}$ K respectively. It is to be noted that a similar calculation within the model (but without the present modifications) was communicated earlier \citep{Jha10}, where we showed the instability window in the high temperature region only. For a better correlation, we investigated the same in the lower temperature region as well, which is compared with the present work in Fig. \ref{omegaplot}. Here the instability window occurs between temperatures $2.2\times10^{5}$ K and $5\times10^{10}$ K while the minima occurs at $0.038$ $\Omega_K$, which corresponds to a temperature of $5.1 \times 10^{9}$ K. 
The model modification has brought down the value of the nuclear incompressibility from 300 MeV to 211 MeV and the symmetry energy slope parameter $L$, from 90 MeV to 61 MeV. As mentioned earlier, these values are in very good agreement with data obtained from both terrestrial experiments and astrophysical observations.}

If the rotational frequency of the star exceeds the critical value `$\Omega_C$', then the star will be subject to $r$-mode driven instability. The range in which most observed low-mass X-ray binaries (LMXBs) are found is also indicated, corresponding to a temperature range of 2 $\times$ $10^7$ - 3 $\times$ $10^8$ K and rotational frequencies in the range of 300 - 700 Hz \citep{Chang2004}. LMXBs are found to be outside our instability windows, suggesting that the $r$-mode driven instabilities are suppressed.  

Using a model described in \citep{Shapiro:1983du}, we also calculate the cooling times due to modified Urca processes. We find that the time it takes for the temperature to drop to $\sim 2\times 10^{10}$ K is lower than the gravitational radiation timescale obtained in our case. Hence the instability does not have enough time to develop. However, further cooling to $\sim 2\times 10^{9}$ K takes a few days, allowing enough time for the $r$-modes to grow unstable. Also, if direct Urca processes are allowed to act, the cooling time to a few times $10^9$ K reduces to about a second \citep{Page2000}.  

Further, we use a phenomenological model described in \cite{Owen98} to calculate the spin-evolution of a neutron star. The parameters of this model are the $r$-mode amplitude $\alpha$ and the angular velocity $\Omega$. We assume the star cools through modified Urca processes. Assuming a small initial amplitude of $\alpha_i = 10^{-5}$, we find that a star initially rotating at Kepler frequency $\Omega_K$ with core temperature $10^{11}$ K spins down to $0.3\Omega_K$ while crossing the first window; after which the amplitude drops to zero. The spin is hence stabilized and the star doesn't spin down any further as it crosses the second window. We repeat the calculation for initial amplitudes between $10^{-3}$ and $10^{-7}$ and find that the final spin period remains between $0.29-0.31~\Omega_K$. The evolution curve corresponding to $\alpha_i = 10^{-5}$ is plotted in Fig. \ref{omegaplot}.

\section{\label{sec:summary}Summary}
A model of a rotating neutron star with a hyperon core is considered in the present study using an effective model calibrated to the symmetry energy and its slope parameters to the data obtained from terrestrial experiments as well as astrophysical observations. With appropriately tuned hyperon couplings to the respective potentials, the mass-radius profile is obtained. We get a mass of $2.56 M_{\odot}$ for the pulsar rotating at its Kepler frequency $\Omega_K = 1420$ Hz, which is an increase of $\approx 25\%$ over the static case $(\Omega = 0)$. Consequently, the equatorial radius is found to be 16.5 km, which is nearly 40\% larger in comparison to the static case. The obtained mass is in agreement with that of the secondary companion of the GW190814 event, as well as many other observed high mass pulsars. The results are also in excellent agreement with the compactness parameter value obtained from the Bayesian analysis of the pulsar $PSR~J0030+0451$. The coefficient of bulk viscosity arising from the hyperon rich neutron star matter is calculated and the resultant damping of the $r$-mode instability is also studied. Presence of two instability windows is seen in the analysis. 
We report that hyperonic bulk viscosity completely damps the growth of $r$-modes between $\sim 10^8$ K to $\sim 10^{9}$ K, creating the stable region in between the windows. 
Also, at temperatures below $6.9 \times 10^4$ K, shear viscosity suppresses the instability, while the stability above $8.44 \times 10^{10}$ K is due to Urca processes.
We compare the results of the present model to that of \cite{Jha10}, where $\sigma-\rho$ cross-coupling was not considered. We find that there is a significant reduction in the region of instability in our model due to hyperon bulk viscosity. The minima of the instability window obtained in \cite{Jha10} has increased by nearly 4 times to $\sim 0.15\Omega_K$ in the present model, indicating a much more effective damping.  
The appearance and concentrations of particles, particularly the $\Lambda^0$ and $\Sigma^-$ hyperons are different for both equations of state. In the present case, only $\Lambda^0$ hyperons are present up to a radius of 4.38 km, beyond which $\Sigma^-$ hyperons appear and then disappear again at a radius of 9.16 km. In the previous model, both these particles were present in approximately similar concentrations at higher densities. The spin evolution of a neutron star is studied for the present model and it is found that the instability can reduce the angular velocity up to $\sim~$0.3~$\Omega_K$.
It will be interesting to look into the $r$-mode suppression in the presence of superfluidity or magnetic field within this model.
\section{Data availability} The data presented in this study are available on request from the corresponding author.
\section{Acknowledgment}
We would like to thank the anonymous referee
whose comments/suggestions have improved the quality of the manuscript significantly.
\bibliographystyle{mnras}
\bibliography{hyp}


\bsp 
\label{lastpage}
\end{document}